\documentclass[proof]{WileyASNA-v2}
\articletype{Proceeding}%
\usepackage{graphicx}
\graphicspath{ {./images/} }
\usepackage{ulem}
\usepackage[innercaption]{sidecap}
\usepackage{xcolor}
\definecolor{darkred}{cmyk}{0,1,1,0.45}

\begin{document}

\title{Destination exoplanet: Habitability conditions influenced by stellar winds properties}

\author[1,2]{Judy J. Chebly*}

\author[1]{Juli\'{a}n D. Alvarado-G\'{o}mez }

\author[1,2]{Katja Poppenhaeger}

\authormark{JUDY J. C. \textsc{et al}}

\address[1]{\orgname{Leibniz Institute for Astrophysics}, \orgaddress{\state{14482 Potsdam}, \country{Germany}}}

\address[2]{\orgdiv{Institute of Physics and Astronomy}, \orgname{University of Potsdam}, \orgaddress{\state{14476 Potsdam-Golm}, \country{Germany}}}

\corres{*Judy J. Chebly, \email{jchebly@aip.de}}

\presentaddress{Leibniz Institute for Astrophysics Potsdam An der Sternwarte 16, 14482 Potsdam, Germany.}

\abstract{The cumulative effect of the magnetized stellar winds on exoplanets dominates over other forms of star-planet interactions. When combined with photoevaporation, these winds will lead to atmospheric erosion. 
This is directly connected with the concept of Habitable Zone (HZ) planets around late-type stars. 
Our knowledge of these magnetized winds is limited, making numerical models useful tools to explore them. In this preliminary study we focus on solar-like stars exploring how different stellar wind properties scale with one another. We used one of the most detailed physics-based models, the 3D Alfv\'{e}n Wave Solar Model part of the Space Weather Modeling Framework, and applied it to the stellar winds domain. Our simulations showed that the magnetic field topology on the star surface plays a fundamental role in shaping the different stellar wind properties (wind speed, mass loss rate, angular momentum loss rate). We conclude that a characterization of the Alfv\'{e}n surface is crucial when studying star-planet interaction as it can serve as an inner-boundary of the HZ.}

\keywords{Stellar winds, Exoplanet, Mass loss, Angular momentum loss, Alfv\'{e}n surface, Space weather}

\maketitle

\section{Introduction}\label{sec1}

The majority of what we know about stellar winds comes from our knowledge of the Sun. We believe that the same mechanisms behind the coronal heating and stellar wind acceleration on the Sun, take place on solar-like stars and low-mass main sequence stars. All of these objects are X-ray sources with coronal temperatures of several million Kelvin. The high gas-pressure gradient in the hot plasma that surrounds these stars will expand as supersonic wind (Parker \citeyear{Parker1958}). 

However, there are large uncertainties in our knowledge of the evolution of stellar winds on the main sequence, due to a lack of direct measurements as well as an incomplete understanding of the solar wind. This reflects in our ability on using stellar winds for rotational evolution models, particularly as some stars experience a rapid spin-down whereas others of the same mass and age do not (Spada et al. \citeyear{Spada2011}). 

Given the sensitivity of planetary atmospheres to stellar wind and radiation conditions, these uncertainties can be significant for our understanding of the evolution of planetary environments. Since the information is so limited, models provide a pathway to explore how these stellar winds look like, and how they behave. 

A detailed parametrization of these magnetized winds is crucial especially when it comes to close-in systems such as M dwarfs where the HZ is close to the star. Current observational constraints  indicate that M dwarfs have winds comparable or even stronger than that of a G star (Wood et al. \citeyear{Wood2021}). As the wind strength experienced by a planet decreases with the square of its orbital distance, even relatively weak winds will have an extremely strong effect on close-in exoplanets.

\begin{figure}[h!]
\vspace{-0.6cm}
\centerline{\includegraphics[width=0.5\textwidth]{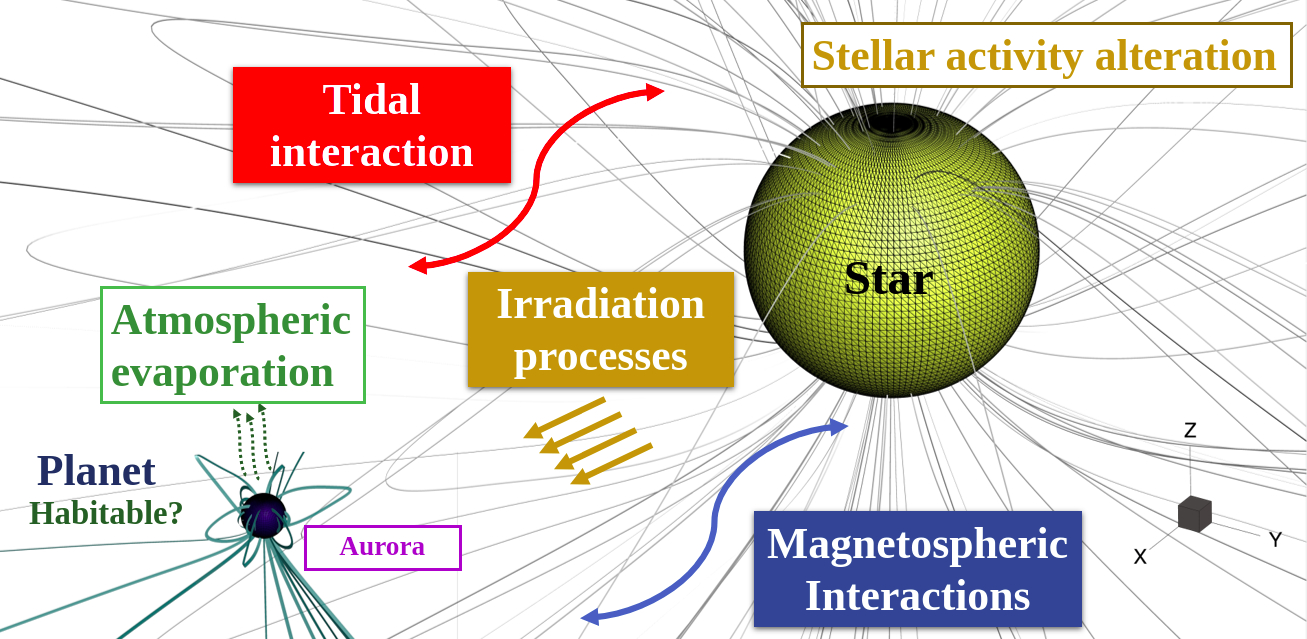}}
\caption{Schematic view of the different star-planet interactions and some of their expected influence.}\label{fig1}
\vspace{-0.6cm}
\end{figure}

In addition, as illustrated in Fig.~\ref{fig1}, when a planet roams in the vicinity of its star, different interactions can occur such as tidal (Ibgui et al. \citeyear{Ibgui2010}; Winn et al. \citeyear{Winn2010}), magnetic (Lanza et al. \citeyear{Lanza2013}, Cohen et al. \citeyear{Cohen2018}), and processes mediated by radiation (Bourrier et al. \citeyear{Bourrier2013}; Fossati et al. \citeyear{Fossati_2013}). 

The magnetic perturbations may lead to a modification of the stellar activity (Shkolnik et al. \citeyear{Shkolnik2008}), but may also be observationally swamped by intrinsic stellar variability (Poppenhaeger et al. \citeyear{Poppenhaeger2011}). They can potentially induce planetary aurorae and associated radio emission (Zarka et al. \citeyear {Zarka2007}; Grie{\ss}meier et al. \citeyear {Griemeier2007}). Moreover, the radiation emitted by the star provokes planet inflation, yielding the outer material more easily taken away by the stellar wind, leading to atmospheric evaporation on the planet (Gronoff et al. \citeyear {Gronoff_2020}).

While magnetic activity decreases with time reducing the quiescent and transient phenomena such as flares and high-energy emission (e.g., ~Skumanich ~\citeyear{Skumanich1972}), stellar winds persist throughout the entire stellar evolution (Wood et al. \citeyear{Wood2004}).
For this reason, their cumulative effect will be dominant for both, the star's angular momentum evolution (Garraffo et al. \citeyear{Garraffo2015}), as well as for possible exoplanets orbiting in the system, affecting in this way the expected  habitability conditions (Alvarado-Gómez et al. \citeyear{2016A&A...594A..95A}, Meadows et al. \citeyear{Meadows2018}). This is because sufficiently strong magnetized winds can erode a stable atmosphere and render a planet inhabitable (Khodachenko et al. \citeyear{Khodachenko2007...7..167K}; Zendejas et al. \citeyear{Zendejas2010}; Vidotto et al.~\citeyear{Vidotto2011}). 

Here we study how different properties of the magnetized stellar wind scale with one another and isolate the most important dependencies between the parameters involved. These results will be later used to constrain planet habitability from a stellar wind perspective. 

Section \ref{sec2} contains a description of the numerical model and parameters employed in the stellar wind simulations. Preliminary results, including dependencies with stellar rotation period, magnetic field strength, and geometry, are presented in Sect.~\ref{sec3}. A discussion of these results is presented in Sect.~\ref{sec4} and we conclude our work in Sect.~\ref{sec5}.

\vspace{-0.2cm}
\section{Model description}\label{sec2}

In this work, we employ the Alfv\'{e}n Wave Solar Model (AWSoM-R, Van der Holst et al. \citeyear{vanderHolst2014}, Sokolov et al. \citeyear{Sokolov2021}), which is part of the Space Weather Modeling Framework (SWMF, T{\'o}th et al. \citeyear{Toth2012}). This model has been tested, validated, and was proved to be one of the closest models in agreement with observations (Sachdeva et al. \citeyear{Sachdeva2019}). The model uses the BATS-R-US code to solve in 3D the MHD equations for the conservation of mass, momentum, magnetic induction, and energy in finite volume form (Powell et al. \citeyear{Powell1999}). 

Simulations employing the SWMF can cover a domain extending from the solar chromosphere up to 1 au in heliosphere and beyond (T{\'o}th et al. \citeyear{Toth2005}). The distribution of the surface magnetic field of the star is the main driver of the solution, from which the heating of the corona and the stellar wind acceleration are calculated self-consistently.

Our main assumption is that the mechanism incorporated in the AWSoM-R model can be extended to stars of different spectral type than the Sun. This is justified due to the fact that the coronal emission from these stars shows several similarities with that of the Sun (Testa et al. \citeyear{Testa2015}). In order to use the model on solar-like stars and lower main sequence stars, we modify the parameters that control the properties of the winds (e.g. magnetic field strength, geometry, stellar rotation, stellar mass, etc). In this way, we create a generic synthetic grid of simplified models from which we can study and extract the dependencies of interest.

We employ the Solar Corona (SC) module of the SWMF whose 3D domain encompasses the region between the surface of the star ($\sim1~R_{\bigstar}$) up to $85~R_{\bigstar}$. We use a spherical grid with a maximum base resolution of $\Delta R = 0.025~R_{\bigstar}$, $\Delta \Phi = 1.4^{\circ}$, stretching radially following a logarithmic function of $R$. The choice of the grid resolution affects the speed of the convergence of the simulation to a steady-state stellar wind solution. In all our models we consider a restricted case in which the magnetic field axis is perfectly aligned with the rotation axis of the star and all the initial solar parameters of AWSoM-R are kept intact except the ones related to the surface magnetic field distribution as discussed below.

In the following sections, we will talk about each stellar wind parameter separately showing how different parameters changes their behavior.

\vspace{-0.2cm}
\section{Results}\label{sec3}

\subsection{Average terminal stellar wind speed}

We calculated the average terminal wind speed by integrating over 3 different spherical shells in the vicinity of the outer boundary of the simulation ({70} $R_{\bigstar}$, {75} $R_{\bigstar}$, {85} $R_{\bigstar}$).

\begin{figure}[h!]
\centerline{\includegraphics[width=0.45\textwidth]{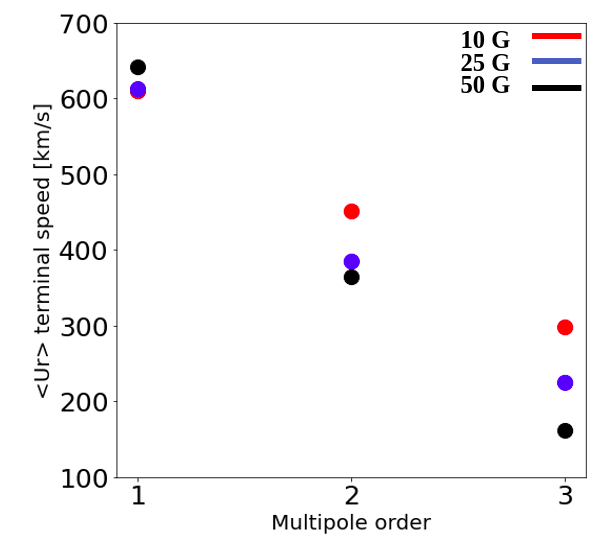}}
\caption{Average terminal stellar wind speed as a function of the surface magnetic field properties (strength and geometry) for a fixed stellar rotation ($P_{rot} = 30$~d). The stellar wind emerging from the simpler geometries is faster.
}\label{fig2}
\vspace{-0.5cm}
\end{figure}

Our calculation of the average terminal radial wind speed $\left<U_{r}\right>$ showed that the more complex the surface geometry, the slower the wind speed becomes (Fig.~\ref{fig2}). For a dipolar field (multipole order of 1) the wind speed is $\sim 600$ km/s almost two times faster than the one of the octopole (multipole order 3). This result is expected since simpler geometries have a smaller number of closed field lines, allowing the plasma to flow out easier (analogous to the solar wind during activity minimum).

We can also see that the distribution of the wind speed changes when we go from a dipole to higher orders. The 10~G wind speed becomes faster than the 50~G, and this is due to the modification in the plasma density when we change the magnetic field geometry (Vidotto \citeyear{Vidotto2021}).

Moreover, Fig.~\ref{fig2} shows that the magnetic field strength plays a secondary role in modifying the wind speed compared to changes in its global geometry. For instance, the simulations with a dipole field of $B_{r} = 10$~G and $B_{r} = 50$~G, yield stellar wind speeds of 609 km/s and 642 km/s, respectively.
Still, the increase in magnetic field strength will modify the distribution of both, the stellar wind speed and the coronal density. The combination of these two dictates the mass loss rate as discussed in the following section. 

We also notice that when the geometry becomes more complex the spread in the terminal wind speed between the different magnetic field strengths becomes wider. Because the more complex magnetic field structure will introduce more streamers, those will carry away higher densities especially when we go higher in the magnetic field strength. The increase in streamers will reduce the local wind speed.

\subsection{Mass loss rate}

Stellar mass loss through magnetized winds is an important parameter to consider when trying to understand the rotational evolution of a star, and it is possibly one of the most important drivers of atmospheric evaporation of exoplanets.

The Alfv\'{e}n surface (AS) sets the boundary between these winds and magnetically-coupled outflows that do not carry angular momentum away from the star (such as loops and prominences). It is a mix between wind speed, plasma density, and the local magnetic field strength. From a numerical approach, this surface is used for different stellar wind parameters calculations. Formally, the AS is given when the Alfv\'{e}nic Mach number --calculated as the ratio of the stellar wind speed to the Alfv\'{e}n speed-- is equal to one ($M_{a}$ = 1). 

\begin{SCfigure}[1.0][!h]
\includegraphics[width=0.375\textwidth]{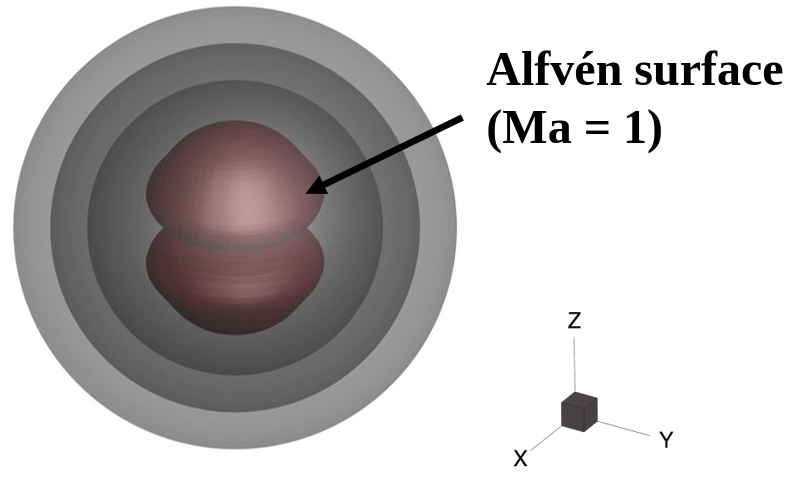}
\caption{Illustration of the numerical calculation of the mass loss rate.}\label{fig3}
\vspace{-0.5cm}
\end{SCfigure}

When calculating the stellar mass loss rate we consider 3 spherical shells located beyond the AS (Fig.~\ref{fig3}) to integrate the mass flux. This is because beyond the AS all the outflows are considered to be carrying mass that will be lost with the wind.\vspace{-0.5cm}

Based on the continuity equation in MHD, the mass should be conserved everywhere, so that a similar mass loss rate is obtained from the integration across each sphere:

\begin{equation}
\mathcal (\frac {dM} {dt}) = \rho (\textbf{u} \cdot dA)\label{eq3}
\end{equation}

Here, $\rho$ represents the wind density, \textbf{u} is the wind speed vector and $dA$ is the vector surface element.

\begin{figure*}[t]
\centering
\includegraphics[width=0.45\textwidth]{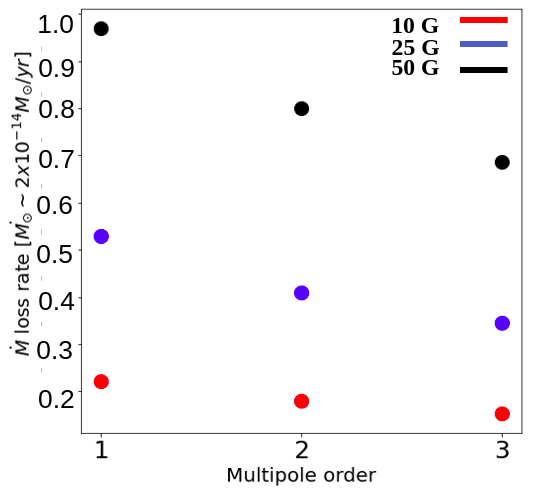} \includegraphics[width=0.45\textwidth]{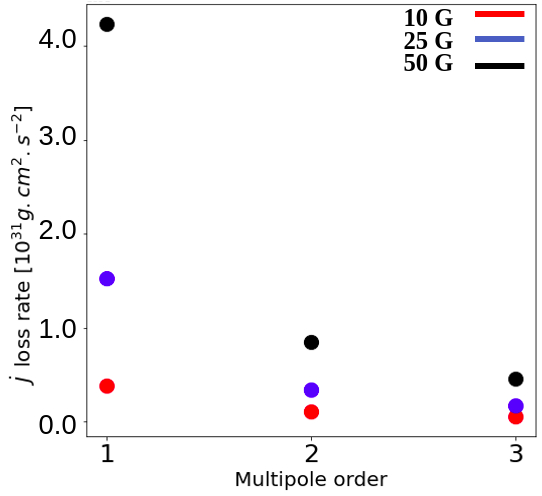}
\caption{Mass loss rate (left) and angular momentum loss rate (right) as a function of the surface magnetic field properties (strength and geometry) for a fixed stellar rotation ($P_{rot} = 30$~d). The mass loss rate is highly affected by the magnetic field strength, while the angular momentum loss rate is higher for simpler geometries and stronger magnetic field strength.}\label{fig4}
\vspace{-0.4cm}
\end{figure*}

We expect that more active stars should in principle have more mass loss. They have strong magnetic fields, so they should have more energy to provide to these winds. The mass loss rate also decreases for higher multipole orders, in line with the argument of Garraffo et al. (\citeyear{Garraffo2015}) which confirms that simpler geometries power up more mass loss (Fig.~\ref{fig4}, left).

\vspace{-0.2cm}
\subsection{Angular momentum loss rate}

For the numerical calculation of the angular momentum loss rate, we integrated every single point on the AS. We took into account its actual shape rather than considering it as a sphere.
The relation for calculating the angular momentum loss in spherical coordinates is provided by
(see Fig.\ref{fig4}).
\begin{equation}
\mathcal (\frac {dJ} {dt}) = \Omega\rho R^{2} \sin^{2} \theta  (\textbf{u} \cdot dA)\label{eq4}
\end{equation}

Here $\Omega$ is the angular frequency of the star and $\theta$ is the angle between the lever arm and the rotation axis. 
$\Omega$ changes with the different stellar rotation $\Omega = 2\pi/P_{rot}$, hence it is not the same in all realisations of the simulations.

\begin{figure*}[t]
\centerline{\includegraphics[width=0.9\textwidth]{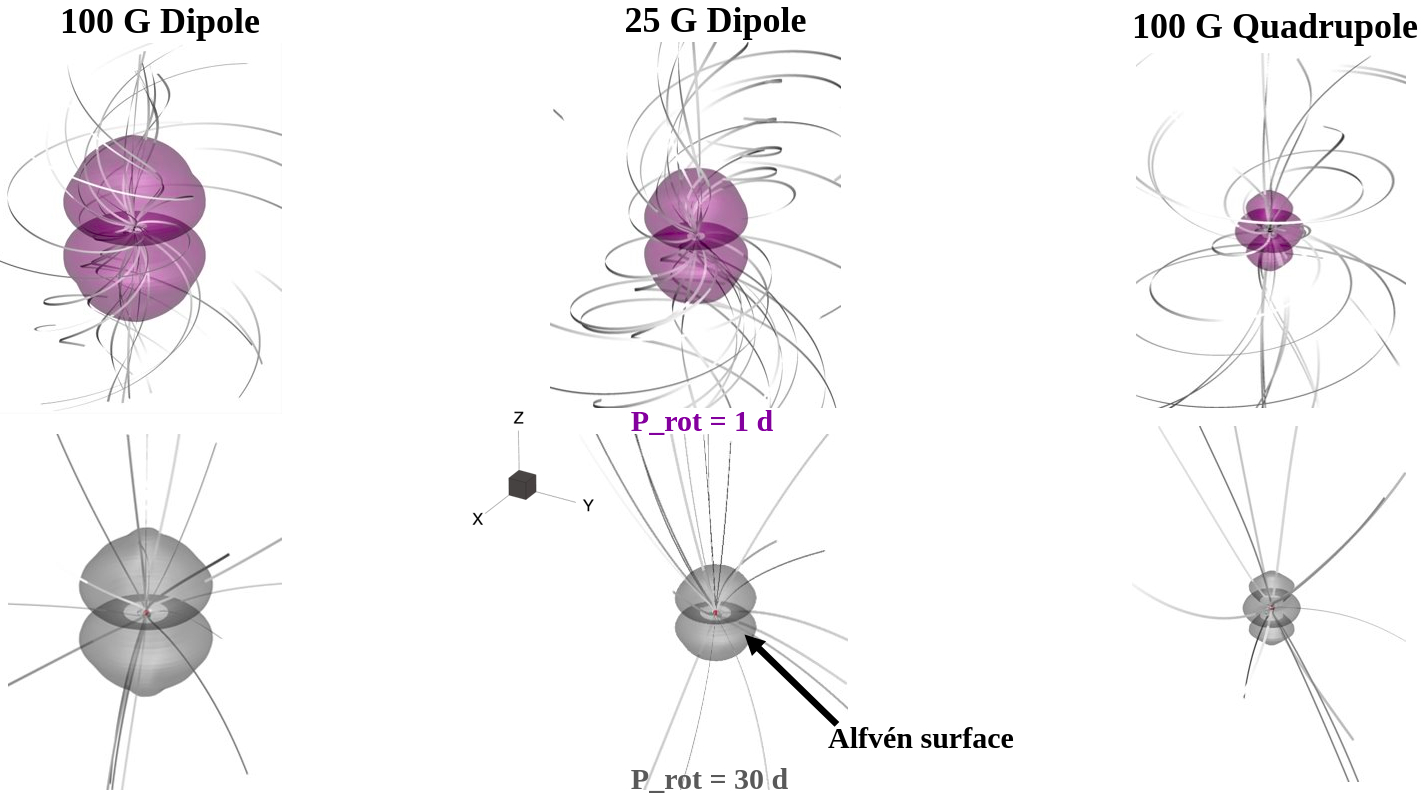}}
\caption{Illustration of the effect of magnetic field strength and geometry on the AS for different values of the stellar rotation period (\textit{Top:} $P_{rot} = 1$~d; \textit{Bottom:} $P_{rot} = 30$~d. The magnetic fields anchored to the star will exert a drag force over the stellar surface. A fast rotation will lead to winding up the magnetic field which will be tighter as we increase the rotation. The AS is highly affected by the change in the surface geometry. The field of view of the visualizations is $243.7~R_{\bigstar}$.\label{fig5}}
\vspace{-0.4cm}
\end{figure*}

For cool main sequence stars, the magnetised wind is believed to carry away stellar angular momentum, causing decay in both their rotation rates and their high-energy radiation. These losses will influence the atmosphere of any planets in the system (Johnstone~et~al.~\citeyear{Johnstone2021}).

The dipole configuration is the one that has the most effective angular momentum loss. What Fig.~\ref{fig4} (right) is showing is basically the ballerina effect: the star is spinning fast when we have more complex geometries.
As we go to higher multipole orders, the number of closed field lines will increase, reducing the AS size and therefore the angular momentum loss rate.

We also see that a stronger magnetic field has a higher angular momentum loss rate. A $50$~G magnetic field will lead to an angular momentum loss rate larger than the $25$~G but the difference between these cases is up to a factor of $\sim 1.5$, while the difference from a $50$~G dipole to a quadrupole is a factor of $\sim 3$ (Fig.~\ref{fig4}, right). There is a large increase by almost an order of magnitude of the angular momentum loss just by changing the geometry from a quadrupole to a dipole for the same magnetic field strength. In Fig.~\ref{fig5} we can see that for the same magnetic field strength, simpler geometries have larger AS and the surface decreases whenever we have higher multipole orders.

We can say that the mass loss contributes to the angular momentum loss rate, but here the quantity that is playing a major role is the size of the AS. This result was expected since the angular momentum loss rate is dependent on the AS size ($R^2$ dependency in Eq.~\ref{eq4}).

\vspace{-0.2cm}
\section{Discussion}\label{sec4}
The redistribution of the plasma density due to the change in the magnetic field geometry plays an important role in the behavior of the different stellar wind properties. The drastic change in the behavior of the wind property was clearly seen when we calculated the terminal wind speed. This means that when trying to understand what is the role played by the winds and by the environment they generate for the planets is much more complicated than simply knowing what is the magnetic field strength of the star. 
The effect of the geometry on the terminal stellar wind speed, mass loss, and angular momentum loss rate was highlighted in Sect.~\ref{sec3}.

Stellar activity cycles change globally the geometry of the magnetic field. For example, the Sun during the solar cycle changes from having dipole properties to a more quadrupole configuration, with wind speeds of $\sim$800 km/s and $\sim$300 km/s, respectively. The maximum solar wind speed is higher than the one of the 10~G dipole (609 km/s) of {$P_{rot} = 30$~d} which means that the solar wind speed is lower in density. The plasma becomes denser as we go to higher multipole orders. 
Moreover, the velocity and plasma density dictates the dynamic pressure of the wind, which is a crucial parameter to explore when studying star-planet interaction.
These parameters are connected to the AS and will change the surface size and modify the stellar wind properties.

The size of the AS surface determines both the total area of integration, as well as the size of the lever arm that applies a torque on a star to spin it down.  Equations \ref{eq3} and \ref{eq4} show the interplay between the actual mass flux through the AS (dictated by the density, and magnetic field geometry) and its size. 
The magnetic particles carried away by this loss can erode the atmosphere of the planet and transform a habitable planet into a non-habitable one even if located in the habitable zone (Lammer et al. \citeyear{Lammer2012}; Vidotto et al. \citeyear{Vidotto2011}).

It has been shown that the angular momentum loss controls the bimodal distribution of rotation periods that we observe in young open clusters (Skumamich et al. \citeyear{Skumanich1972}). 
Brown (\citeyear{Brown2014}) explained this distribution based on the idea that rotating stars fall into two different regimes; one in which the dynamo is strongly coupled to the wind and another one where it is weakly coupled to the wind.
The initial distribution of rotation periods will yield to another one when the star becomes older. This is manifested as a bipolarity in how the rotation periods distribute (see Garraffo et al. \citeyear{Garraffo2018}).

In Fig.~\ref{fig4} (right) we have an estimation of the angular momentum loss rate for stars with the same characteristics, but different rotations. We notice that fast-rotating stars have a large angular momentum loss rate, because they have big AS. These stars will spin down faster than the slow rotators.

Angular momentum loss rates are highly affected by the change of geometry rather than by the change of magnetic field strength. In the study of Finley et al. (\citeyear{Finley2018}) they show that angular momentum loss rates vary by 30-40\% over the solar cycle. This shows that despite the dramatic change in the surface magnetic field strength of the Sun during the cycle, the changes in the solar wind properties are not drastic. Additional works have been done showing the angular momentum changes in function of the star rotation period (e.g. Bouvier et al. \citeyear{Bouvier2014};
Shoda et al. \citeyear{Shoda2020}).

\vspace{-0.4cm}
\section{Conclusion}\label{sec5}
The magnetic field geometry of the star surface plays a fundamental role in defining the AS shape and size leading to alteration in stellar wind properties. Hence, a detailed characterisation of the AS is needed because it sets the boundary between the stellar winds and magnetically-coupled outflows and especially the equatorial AS - where many of the known exoplanets are located (e.g. Winn et al. \citeyear{Winn2015}).

The AS can be used as the inner boundary of the actual HZ. However, we may find that the inner boundary that was once predicted by the classical description based on surface temperature, is pushed further away from the star, because atmospheres may be easily eroded inside the HZ.

If the planet is almost embedded in the magnetic field of the star we can imagine that we have reconnection events taking place. If the planet is magnetized, we can have interactions between the magnetic fields. 

Simulating the stellar wind domain requires prior knowledge of the large scale surface star magnetic field geometry which can be retrieved to some extent by techniques such as Zeeman Doppler-Imaging (ZDI). It has been argued that ZDI reconstructions have limited spatial resolution because they are insensitive to the small-scale surface field. However, far away from the star, these small-scale phenomena on the star surface won't be very relevant but they will matter when we want to simulate the corona (Garraffo et al. \citeyear{Garraffo2013}, Alvarado-G\'{o}mez et al.~\citeyear{2016A&A...588A..28A}). The more constraints we have on stellar properties and the magnetic field of the star, like a detailed ZDI maps and also constraints by Zeeman broadening, the more accurate models we will be able to obtain. Different observation satellites e.g. XMM-Newton, Chandra and the future mission ATHENA will help us to improve and generate more realistic numerical simulations.

For further investigation, we will explore different main-sequence spectral types with the ultimate goal of constraining planetary habitability from a stellar winds perspective. 

\vspace{-0.4cm}
\section*{Acknowledgments}

This research is supported by the \fundingAgency{German Leibniz
Community} grant \fundingNumber{P67/2018}.

\bibliography{Wiley.bib}

\end{document}